\documentclass[aps,twocolumn,superscriptaddress,prb]{revtex4}
\usepackage{graphicx,rotating,subfigure,amsmath,amsfonts,amssymb,delarray,psfrag}
\renewcommand{\vec}[1]{\boldsymbol #1}
\newcommand{\e}{\text{e}}
\newcommand{\im}{\text{i}}

\begin{document}
% You should use BibTeX and apsrev.bst for references
%\bibliographystyle{phys}

%\preprint{}

\title{Ground-state properties of two-dimensional dimerized Heisenberg models}
% Optional argument for running titles on pages
%\title[]{}


% other informatio
\author{J. Sirker}
\email[]{sirker@fkt.physik.uni-dortmund.de}
%\thanks{}
\affiliation{Theoretische Physik I, Universit\"at Dortmund, Otto-Hahn-Str.\!\! 4, D-44221 Dortmund, Germany}

\author{A. Kl\"umper}
\email[]{kluemper@fkt.physik.uni-dortmund.de}
%\thanks{}
\affiliation{Theoretische Physik I, Universit\"at Dortmund, Otto-Hahn-Str.\!\! 4, D-44221 Dortmund, Germany}

\author{K. Hamacher}
\email[]{kontakt@kay-hamacher.de}
%\thanks{}
\affiliation{Theoretische Physik I, Universit\"at Dortmund, Otto-Hahn-Str.\!\! 4, D-44221 Dortmund, Germany}


\date{\today}

\begin{abstract}
The purpose of this paper is to investigate the ground-state properties of two-dimensional Heisenberg models on a square lattice with a given dimerization. Our aim is threefold: First, we want to investigate the dimensional transition from two to one dimension for three models consisting of weakly coupled chains for large dimerizations. Simple scaling arguments show that the interchain coupling is always relevant. The ground states of two of these models therefore have one-dimensional nature only at the decoupling point. The third considered model is more complicated, because it contains additional relevant intrachain couplings leading to a gap as shown by scaling arguments and numerical investigations. Second, we investigate at which point the dimerization destroys the N\'eel ordered ground state of the isotropic model. Within a mapping to a nonlinear sigma-model and linear spinwave theory (LSWT) we conclude that the stability of the N\'eel ordered state depends on the microscopic details of the model. Third, the considered models also can be regarded as effective models for a spin system with spin-phonon coupling. This leads to the question if a spin-Peierls transition, i.e. a gain of total energy due to lattice distortion, is possible. LSWT shows that such a transition is possible under certain conditions leading to a coexistence of long-range order and spin-Peierls dimerization. We also find that the gain of magnetic energy is largest for a stair-like distortion of the lattice.    
\end{abstract}
\pacs{}
%\keywords{}

\maketitle

\section{Introduction}
\label{introduction}
\begin{figure}[!ht]
\includegraphics [width=0.4\columnwidth]{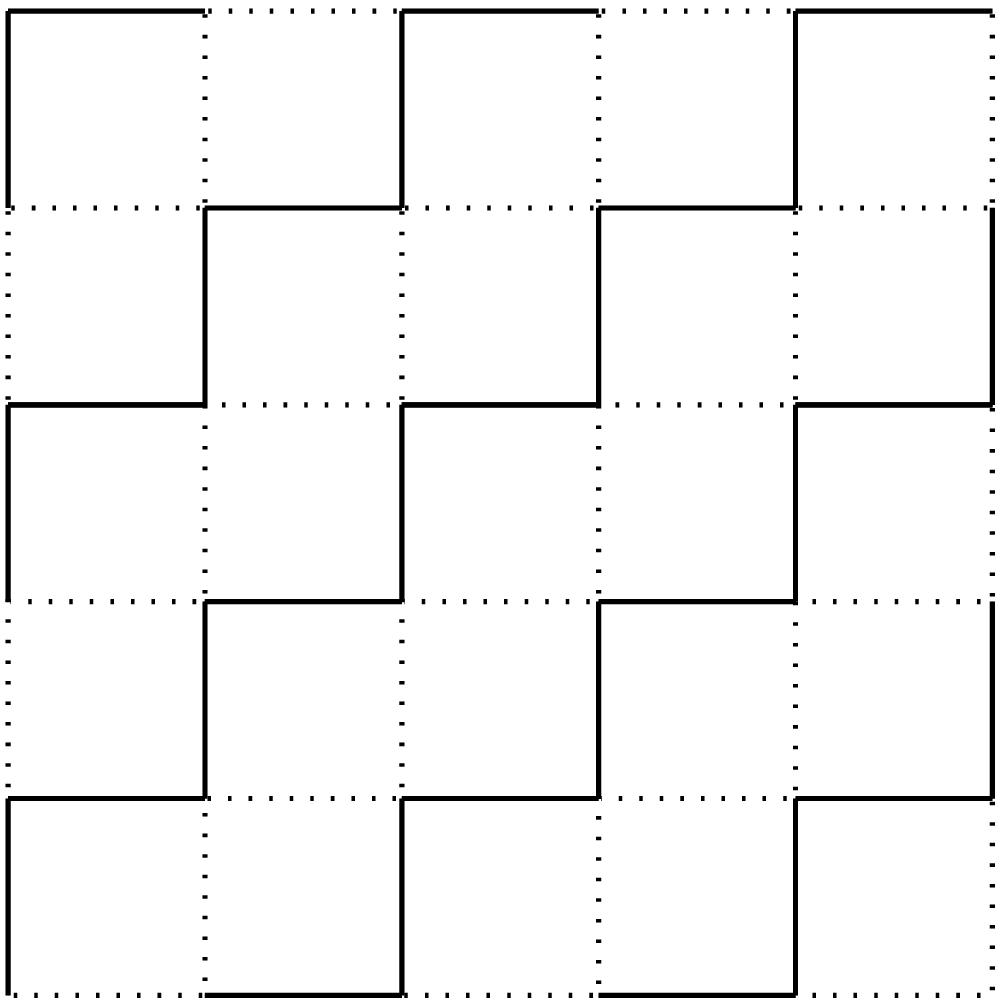}
\caption{(STAIR-MODEL) A thick solid line indicates a strong bond $J(1+\delta)$ and a dashed line a weak bond with strength $J(1-\delta)$. This distortion of the lattice is caused by one transversal phonon with wave vector $(\pi,\pi)$.}
\label{fig1}
\end{figure}

\begin{figure}[!ht]
\includegraphics [width=0.4\columnwidth]{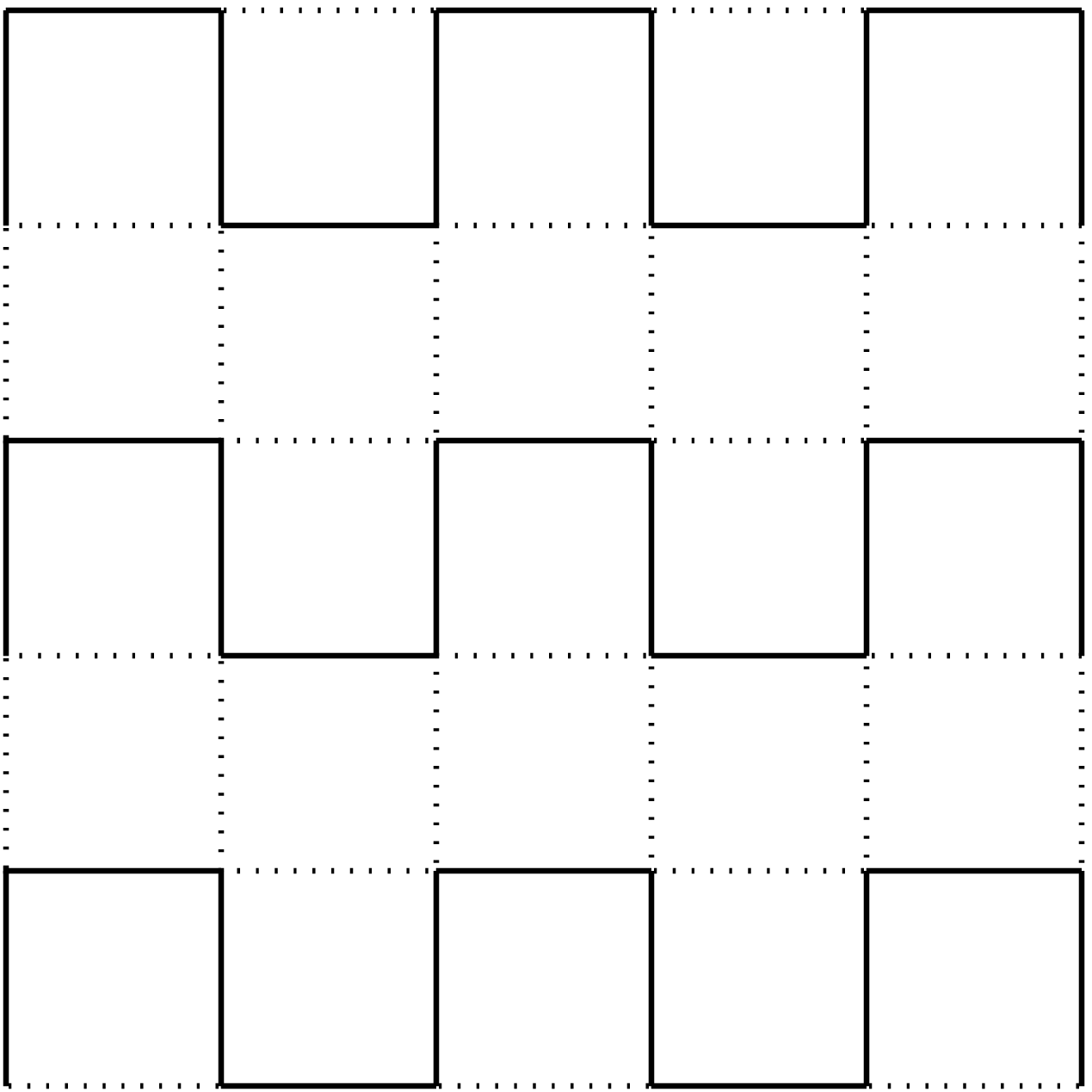}
\caption{(MEANDER-MODEL) Three phonons, a longitudinal and a transversal $(\pi,\pi)$-phonon together with a longitudinal $(0,\pi)$-phonon.}
\label{fig2}
\end{figure}

\begin{figure}[!ht]
\includegraphics [width=0.4\columnwidth]{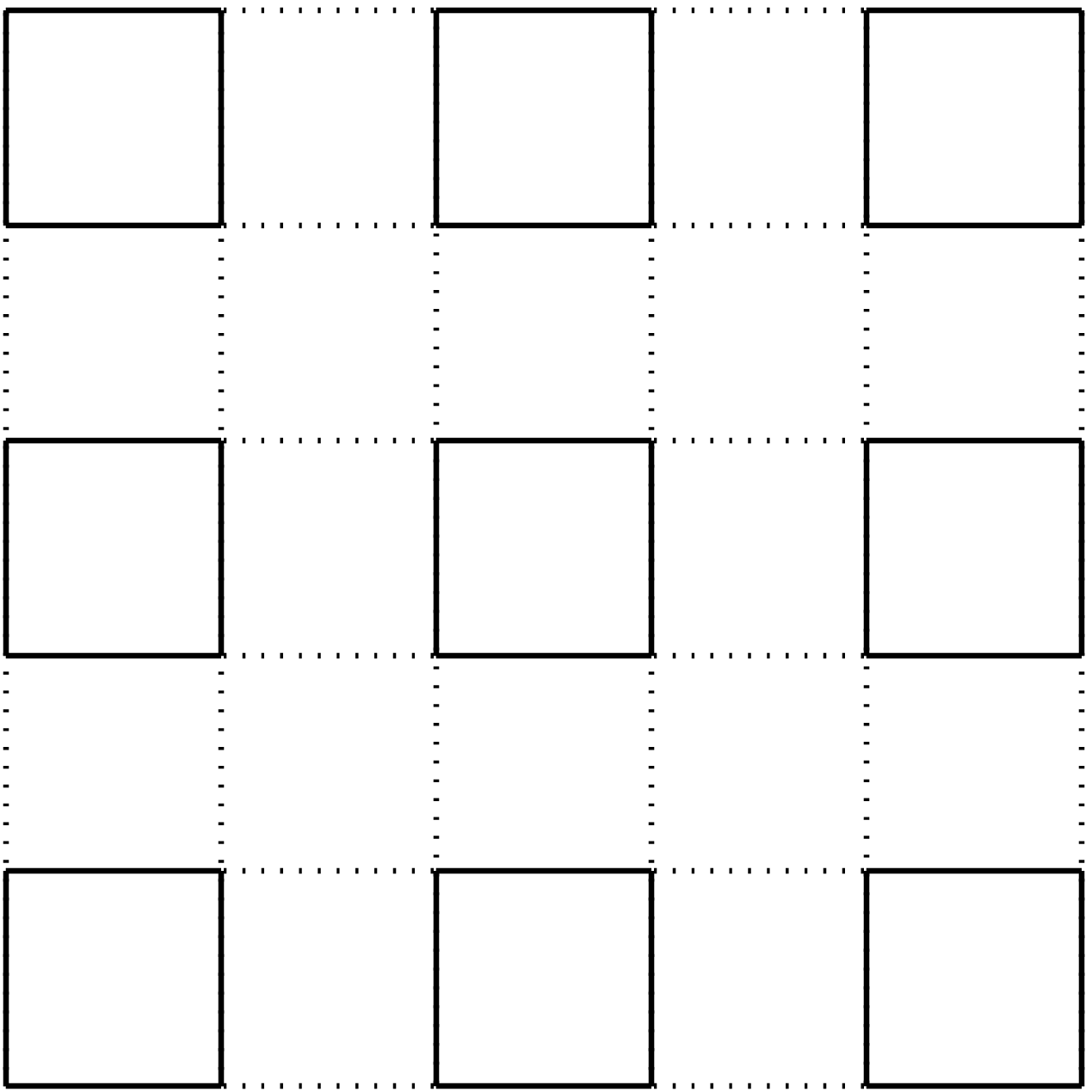}
\caption{(PLAQUETTE-MODEL) Two longitudinal phonons, one with wave vector $(\pi,0)$, the other with $(0,\pi)$.}
\label{fig3}
\end{figure}

\begin{figure}[!ht]
\includegraphics [width=0.4\columnwidth]{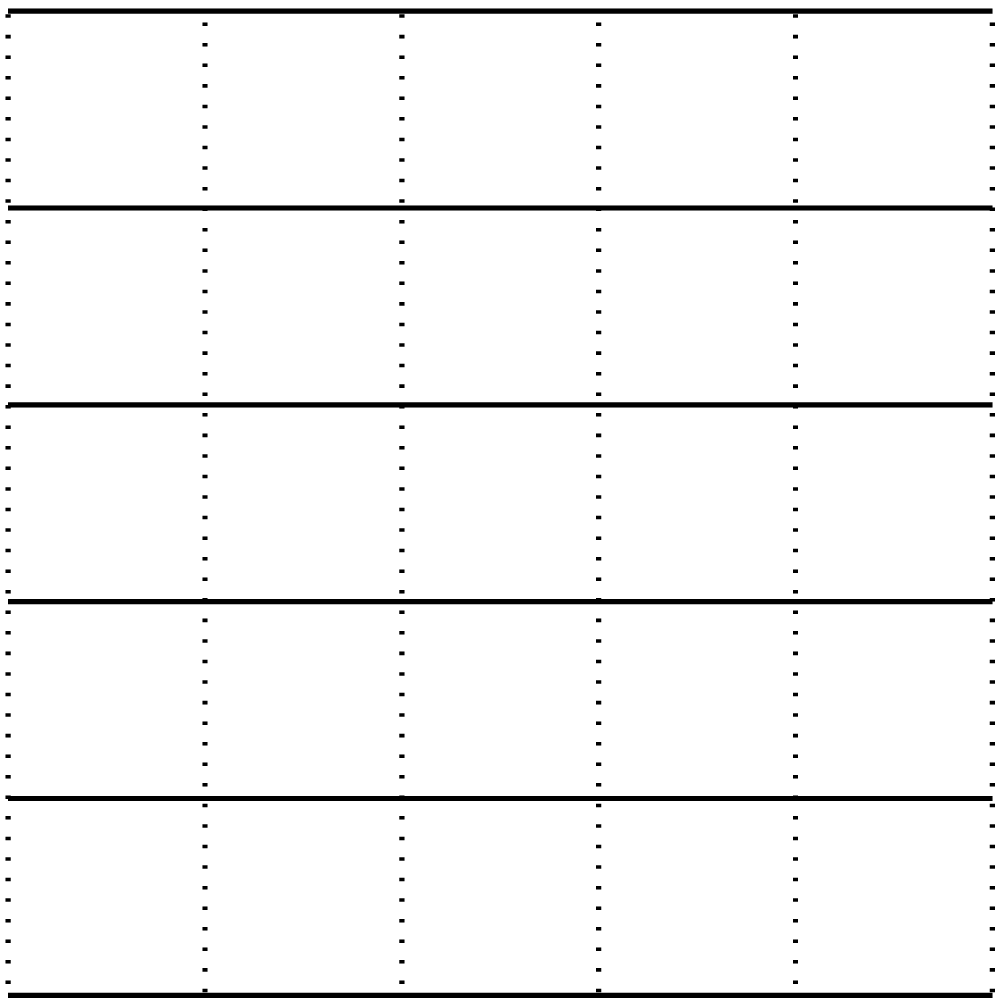}
\caption{($J_x/J_y$-MODEL) Two longitudinal phonons, one with wave vector $(-q,0)$ the other with $(0,+q)$, where $q$ is infinitesimal ($q = 2\pi/\sqrt{N}a$).}
\label{fig4}
\end{figure}

Since many years there is considerable interest, both experimentally and theoretically, in the subject of low-dimensional quantum spin systems, because their properties are strongly affected by quantum fluctuations. The generic model to theoretically study such systems is the well-known Heisenberg model.\\
In one spatial dimension the model with nearest neighbor exchange of spin-1/2 objects, known to be exactly solvable by Bethe ansatz,\cite{Bethe} shows an algebraic decay of its correlation functions at zero temperature and constitutes thereby a quantum critical system. For arbritary spin s, Haldane \cite{Haldane1} has mapped the spin chain onto a nonlinear sigma-model (nl$\sigma$-model) with a topological term for half-integer $s$ and without such a term for integer s. From this result he conjectured that half-integer spin chains are critical whereas integer spin chains have a gap, a scenario that is well established by now. Another interesting aspect of the spin-1/2 Heisenberg chain is its instability towards a structural transition known as spin-Peierls transition.\cite{CrossFisher} \\
Much less is known for the isotropic two-dimensional Heisenberg antiferromagnet with nearest neighbor exchange on a square lattice. Contrary to one dimension no exact solution is available in any limit. LSWT, which does not work in one dimension because of infrared divergencies, is applicable and predicts a N\'eel ordered ground state for the spin-1/2 case, but with a magnetic moment reduced to nearly 50\% of its classical value.\cite{Anderson} This result is also supported, qualitatively and also quantitatively, by numerical work.\cite{BeardBirgeneau,Barnes} For $s\geq 1$ Dyson, Lieb and Simon \cite{DysonLieb} proved a theorem, which shows that the ground state is N\'eel ordered. Like in one dimension it is also possible to map the system onto a nl$\sigma$-model. From a renormalization-group (RG) treatment it is known \cite{Polyakov} that this model exhibits in (2+1)-dimensions a nontrivial critical point $g_c$, which separates a phase with N\'eel-like long-range order ($g<g_c$) from a quantum disordered phase ($g>g_c$) at $T=0$. It has been shown that there is an excellent agreement between theoretical results for this model and experimental measurements on $La_2CuO_4$ in the low-temperature regime if $g < g_c$ is assumed.\cite{ChakravartyHalperin} An interesting problem has been the question, if there is a topological term also in two dimensions, which was finally answered by Haldane,\cite{Haldane2} who concluded that such a term is always absent if the order parameter field is smooth on the scale of the lattice spacing. However, there are tunneling events, which are crucial for an understanding of the disordered phase.\cite{ReadSachdev} \\
\\
In this work we want to consider two-dimensional Heisenberg models on a square lattice with a given alternation of the coupling between nearest neighbor spins. In each spatial direction the coupling should be changing from bond to bond between $J(1+\delta)$ and $J(1-\delta)$ with $J>0$ and $\delta \in [0,1]$ so that the coupling is always antiferromagnetic. There are three topologically different possibilities (cf. figures \ref{fig1}, \ref{fig2}, \ref{fig3}) for arranging such "dimerized chains" on a square lattice if periodicity in each spatial direction is assumed. 
These systems are described by the following Hamiltonian
\begin{eqnarray}
\label{model-Ham}
H &=& J \sum_i \left(1+(-1)^{i(+j)}\delta\right) \vec{S}_{i,j}\vec{S}_{i+1,j} \\
&+& J \sum_j \left(1+(-1)^{j(+i)}\delta\right) \vec{S}_{i,j}\vec{S}_{i,j+1} \; ,\nonumber
\end{eqnarray}
where $\vec{S}_{i,j}$ denotes the spin operator acting on the lattice site $(i,j)$. Choosing both exponents equal to $i+j$ leads to the model shown in figure \ref{fig1}, whereas setting the first one equal to $i$ and the second one to $i+j$ or vice versa leads to the model shown in figure \ref{fig2}. The third considered model (see fig \ref{fig3}) is described by (\ref{model-Ham}) with the first exponent set to $i$ and the second set to $j$.\\ 
The models in figure \ref{fig1} and figure \ref{fig2}, which we want to refer to as stair-model and meander-model, decouple into spin chains at $\delta = 1$. This means that there is a transition from two to one dimension depending on the value of the dimerization $\delta$. Because a model with a coupling $J_x$ in x-direction and a coupling $J_y$ in y-direction (see figure \ref{fig4}) is the simplest model showing such a transition, we want to reexamine this model although it has been studied intensively before. 
The model in figure \ref{fig3} is different from the other three models, because it decouples into plaquettes consisting of four spins for $\delta = 1$.\\
The dimerized models also can be regarded as effective models for a spin system with a spin-phonon coupling treated adiabatically. In such a system the exchange coupling $J$ between nearest neighbors depends linearly on their distance. This is expressed through the deformation parameter of the horizontal bond extending from site $(i,j)$ to the right 
\begin{equation}
\label{spin-phonon}
\delta^h(i,j) = \frac{1}{J}\sum_{\vec{k},s} \lambda_{\vec{k},s} \left( \langle \vec{u}_{i,j}(\vec{k},s) \rangle - \langle \vec{u}_{i+1,j}(\vec{k},s) \rangle \right)_\parallel 
\end{equation}
and the deformation parameter of the vertical bond $\delta^v(i,j)$ being defined analogously. Here $\lambda_{\vec{k},s}$ is the microscopic spin-phonon coupling constant and $\vec{u}_{i,j}(\vec{k},s)$ the local displacement of the atom at the position $(i,j)$ with respect to the phonon wave vector $\vec{k}$ and branch $s$. In the considered models $\delta^h$ and $\delta^v$ are equal to $\pm\delta$. Note that the lattice distortion is static due to the mean-field approach leading to a classical elastic energy. \\

In section \ref{Scaling_arg}, we study the models, which decouple into chains at $\delta = 1$. We address the question whether there is a transition from two- to one-dimensional behaviour already at a $\delta$ smaller than 1 on the basis of some scaling arguments. Because the situation in the meander-model is much more complicated than in the other models due to third-nearest neighbor (NNNN) couplings, we show in section \ref{numerical_inv} numerical results from density-matrix renormalization group (DMRG) and transfer-matrix DMRG (TMRG) to confirm the conclusions drawn from the simple scaling arguments in section \ref{Scaling_arg}. In section \ref{nonlinsigma}, we show that it is possible to map all models onto a 2+1-dimensional nl$\sigma$-model. We use the known RG results for this model to discuss the magnetic properties of the spin models. In section \ref{Linspinwave}, LSWT is applied to support the picture from the RG arguments and to determine a value for the critical dimerization $\delta_c$, where the magnetic order vanishes. Also from this, we get a condition for the spin-Peierls transition and an answer to the question, which structure is energetically prefered. The sections \ref{nonlinsigma} and \ref{Linspinwave} consist of separate subsections for each model. However, in \ref{Linspinwave} we have interchanged the order of the first two subsections in comparison with \ref{nonlinsigma} to keep calculations as simple as possible. In section \ref{Conclusions}, we discuss our results and give some conclusions.   

\section{Scaling arguments}
\label{Scaling_arg}
\begin{figure}[ht]
  \begin{center}
    \leavevmode
    \includegraphics [width=0.9\columnwidth]{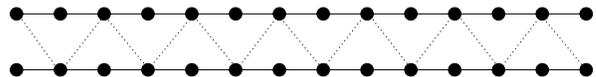}
    \caption{Two chains from the stair-model with the corresponding interchain coupling}
    \label{stair-chains}
  \end{center}
\end{figure}

\begin{figure}[!ht]
  \begin{center}
    \leavevmode
    \includegraphics [width=0.9\columnwidth]{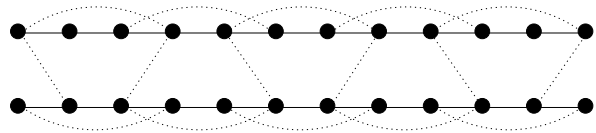}
    \caption{Two smooth chains from the meander-model}
    \label{meander-model}
  \end{center}
\end{figure}

At $\delta=1-\epsilon$ with $\epsilon \ll 1$, the models in figure \ref{fig1}, \ref{fig2}, \ref{fig4} consist of weakly coupled Heisenberg chains. Such a Heisenberg chain with $s=1/2$ is a critical system and the additional weak inter- and intrachain couplings are small perturbations of this critical system. If we pick two such chains from each model with the corresponding interchain coupling proportional to $\epsilon$, we can determine the relevance of the perturbation by calculating the energy-energy correlation function and from this the scaling dimension of the perturbation operator. Let us start with the simplest case, the $J_x/J_y$-model, where $J_y = J(1-\delta)$ and $J_x = J(1+\delta)$. The Hamiltonian of the weak interchain coupling is given by
\begin{equation}
\label{eq:Storham}
 \tilde{H} = J_y\sum_{r} \vec{S}_r^1\vec{S}_r^2 \; ,
\end{equation}
where the upper index labels the two chains. The energy-energy correlation function of this perturbation can be calculated as follows
\begin{eqnarray}
 \langle \sigma_0\sigma_r \rangle_{_0} & = & J_y^2 \sum_{\alpha,\beta} \langle S_0^{\alpha,1}S_0^{\alpha,2}  S_r^{\beta,1}S_r^{\beta,2} \rangle_{_0} \nonumber \\
 % &= & J_y^2 \sum_{\alpha,\beta} \langle S_0^{\alpha,1}S_r^{\beta,1} \rangle \langle S_0^{\alpha,2}S_r^{\beta,2} \rangle \nonumber \\
 &=& J_y^2 \sum_{\alpha} \langle S_0^{\alpha,1}S_r^{\alpha,1} \rangle \langle S_0^{\alpha,2}S_r^{\alpha,2} \rangle \\
&=& \frac{J_y^2}{3} \langle \vec{S}_0\vec{S}_r \rangle^2 = \frac{J_y^2}{3}\cdot\left(\frac{(-1)^r}{r}\right)^2 \; ,\nonumber
\end{eqnarray}
where $\sigma_r=J_y \vec{S}_r^1\vec{S}_r^2$ and $\alpha$, $\beta$ label the components of the spin operator. The subscript "$0$" means calculation in the case of vanishing interchain coupling and in the last relation the known result for the spin correlation function of the homogenous Heisenberg chain is used.\cite{Woynarovich} From conformal field theory it is known that this correlation function decays algebraically like $1/r^{2x}$, where $x$ is the scaling dimension when we disregard multiplicative logarithmic corrections. We therefore conclude that the scaling dimension of the interchain coupling is $x=1$ and represents a relevant perturbation of the critical system. From scaling relations we find, again ignoring logarithmic corrections, the ground-state energy $E_0$ of this system behaving like
\begin{equation}
\label{eq:Grundzustandenergie}
 E_0 \propto J_y^{\frac{d}{d-x}} = J_y^2
\end{equation}
and a gap $\Delta$ is opening with
\begin{equation}
\label{eq:Lucke}
 \Delta \propto |J_y|^{\frac{1}{d-x}} = |J_y| \; ,
\end{equation}
where $d=1+1$ is the dimension of the corresponding classical model. The existence of a gap for the two-leg ladder has also been shown numerically.\cite{WhiteScalapino} In general, there seems to be a gap for an even number of coupled chains, whereas a system with an odd number of chains is gapless. However, it is not possible to determine from scaling arguments if there is a gap or not for an infinite number of arbitrarily weakly coupled chains. Nevertheless, the relevance of the interchain coupling clearly shows that the system scales away from decoupled chains and therefore even at very large, however not perfect dimerization does not behave like decoupled chains. We conclude that the ground state of this model has two-dimensional nature if $J_y \neq 0$ as has been stated before.\cite{Affleck} \\
Also we can pick two chains out of the stair-model and after smoothing the chains, we get the configuration shown in figure \ref{stair-chains}.
Here the interchain coupling is described by the Hamiltonian
\begin{equation}
\label{eq:Storham2}
\tilde{H} = \tilde{J} \sum_{r=1}^{N/2}  \vec{S}_{2r}^1 (\vec{S}_{2r-1}^2+\vec{S}_{2r+1}^2) 
\end{equation}
and by calculating again the energy-energy correlation function, we find that this perturbation also has scaling dimension $x=1$. That leads to the same conclusions as in the $J_x/J_y$-model. \\
The situation is much more complicated in the meander-model, because there is not only an interchain coupling, but also a coupling between third-nearest neighbors within every chain as shown in figure \ref{meander-model}.
First, we want to investigate the intrachain coupling. The operator of this perturbation is given by
\begin{equation}
\label{eq:Storham3}
\tilde{H} = \tilde{J} \sum_{r=0}^{N/2}  \vec{S}_{2r} \vec{S}_{2r+3}
\end{equation}
and we can calculate the corresponding energy-energy correlation function 
\begin{eqnarray}
 \langle \sigma_0\sigma_r \rangle_{_0} & = & \tilde{J}^2  \langle \vec{S}_0 \vec{S}_3  \vec{S}_{2r} \vec{S}_{2r+3} \rangle_{_0} \nonumber \\
 &\propto& \frac{(-1)^{2r}}{2r} \, ,
\end{eqnarray}
where the value $x=1/2$ of the scaling dimension of the singlet
operator $\vec{S}_{2r} \vec{S}_{2r+3}$ has been employed.\cite{Woynarovich}
This means that the NNNN coupling is relevant. By simply applying the scaling relations, we conclude that it destroys criticality and a gap opens with $\Delta \propto |\tilde{J}|^{2/3}$. But there might be some doubt if this scenario is correct, because if we suggest a short range N\'eel order on the critical chain the NNNN coupling is not frustrating. On the other hand, if the NNNN coupling is as strong as the NN coupling, this chain is equivalent to a 2-leg ladder, which does show a gap. We therefore have used the transfer-matrix DMRG and the standard DMRG to test numerically the predictions from scaling. Before we enter the numerical part, we have to analyse the other perturbation in the meander-model caused by the interchain coupling. It turns out that this is again relevant with a scaling dimension $x=1$ as in the other two models. \\
The conclusion from scaling arguments is therefore that the $J_x/J_y$- and the stair-model show one-dimensional behaviour only at the decoupling point. Because in the meander-model the intrachain is more relevant than the interchain coupling, the scaling arguments suggest the existence of a disordered phase between the decoupling point and the phase with two-dimensional antiferromagnetic long-range order. This will be further investigated in sections \ref{numerical_inv} and \ref{nonlinsigma}. 

\section{Numerical investigations}
\label{numerical_inv}
\begin{figure}[!htb]
  \begin{center}
    \leavevmode
    \includegraphics [width=0.9\columnwidth]{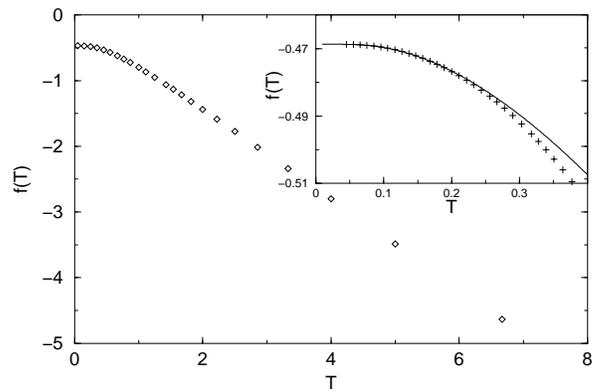}
    \caption{Free energy for the Heisenberg chain with NNNN coupling calculated by TMRG with $J_2/J_1=0.25$, $m=40$ states kept in the DMRG and a Trotter parameter of $\epsilon=\beta/M=0.05$. The inset shows a low-temperature fit with $f \propto e_0 - a T^{3/2} \e^{-\Delta/T}$ and $T\in[0,0.2]$.} 
    \label{TMRG}
  \end{center}
\end{figure}

\begin{figure}[!ht]
  \includegraphics [width=0.9\columnwidth]{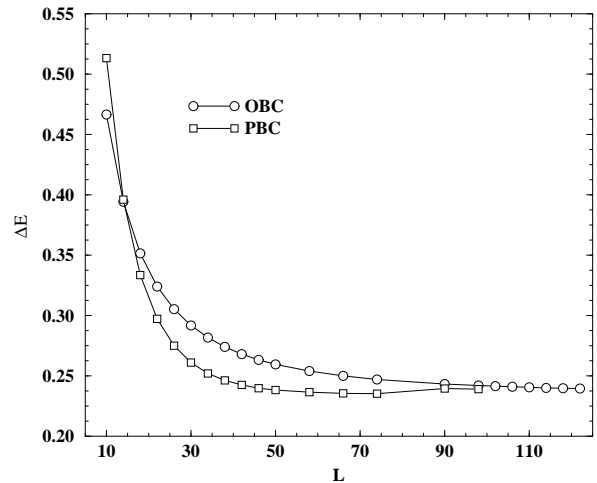}
    \caption{DMRG calculation of the gap $\Delta E$ for finite chains with length $L$ for open and periodic boundary conditions using an extrapolation in the number of states $m$. The lines are guides to the eye.} 
    \label{DMRG}
  \end{figure}

\begin{figure}[!ht]
  \includegraphics [width=0.9\columnwidth]{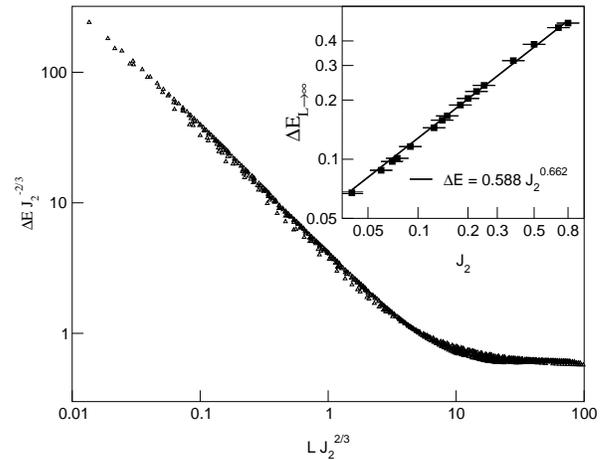}
    \caption{DMRG results for 862 chains with different length $L$ and NNNN coupling $J_2$. Two scaling regimes are visible as discussed in the main body. The inset shows the gap extrapolated to the thermodynamic limit versus coupling $J_2$.}     
	\label{Skalenverhalten}
  \end{figure}

To prove the scaling argument for the NNNN coupling in the meander-model, we have used two numerical methods. The first one is the so called transfer-matrix DMRG (TMRG), which combines White's DMRG idea \cite{WhiteDMRG} with the transfer-matrix approach.\cite{SuzukiInoue} This method has been applied to different quantum chains before \cite{WangXiang,Shibata,Raupach} and yields very accurate results for finite temperature. It is particulary suited, because the thermodynamic limit in quantum space can be performed exactly. Before we state the results, we write down the considered Hamiltonian explicitly:
\begin{equation}
\label{eq:Num1}
H = J_1\sum_{r=0}^{N} \vec{S}_{r} \vec{S}_{r+1} + J_2 \sum_{r=0}^{N/2}  \vec{S}_{2r} \vec{S}_{2r+3}
\end{equation}
The relevant case in this context is a NNNN coupling $J_2$, which is much smaller than the NN coupling $J_1$. \\
The free energy as calculated with the TMRG for $J_2/J_1 = 0.25$ is shown in figure \ref{TMRG}. 
We can now determine if there is a gap or not, because we know from scaling relations that in the low-temperature limit the free energy of a gapless, critical system scales like $f(T) \propto e_0 - a \cdot T^2$ with $e_0$ being the ground-state energy, whereas $f(T) \propto e_0 - a \cdot T^{3/2} \e^{-\Delta/T}$ if there is a gap. We tried to fit a quadratic function to the data and noted that this is impossible, whereas a function as expected for the gapped case fits perfectly with values $e_0=-0.46873 \pm 0.00002$, $a=0.29 \pm 0.03$, $\Delta=0.23 \pm 0.02$ and errors, which are determined by a variation of the fit-region (see inset of fig. \ref{TMRG}). \\
This means that at this strength the NNNN coupling has really destroyed criticality. To test the scaling argument further, we also applied a standard DMRG program to this problem. When using the same parameters $J_2/J_1=0.25$, we find a gap depending on the length of the chain as shown in figure \ref{DMRG}. An extrapolation $L \rightarrow \infty$ then leads to a gap $\Delta_{\mbox{\footnotesize PBC}} = 0.23652 \pm 0.00064$ and a ground-state energy  $e_{\mbox{\footnotesize PBC}} = -0.46841\pm 0.00816$ if periodic boundary conditions (PBC) are applied. For open boundary conditions (OBC), we find a gap $\Delta_{\mbox{\footnotesize OBC}} = 0.23834 \pm 0.00027$ and a ground-state energy $e_{\mbox{\footnotesize OBC}} = -0.46843\pm 0.00003$. Consequently, there is a good agreement between the numerical results from the two different methods. 
Up to now, we have only stated that the system has a gap for one special choice of parameters. Because the infinite chain with $J_2=0$ is a critical system, we can use the RG to study the behaviour of the free energy depending on the length of the chain and of the NNNN coupling $J_2$. In general, we can linearize the RG transformation in the vicinity of a critical Hamiltonian, which is a fixed point of the RG flow, and find that the free energy per lattice site for a classical system scales like $f(g_1,\cdots,g_n) = b^{-ld} f(b^{l \lambda_1}g_1, \cdots,b^{l \lambda_n}g_n)$ if the RG is applied l-times. Here $g_i$ denotes a linear scaling field, $\lambda_i$ is the eigenvalue of the RG transformation and $b$ is the scaling factor. Because the quantum chain with length $L$ is equivalent to a classical system with volume $V=L \times \beta$, with $\beta$ being the inverse temperature, the dimension $d$ is equal to 2 and the relevant scaling fields at $T=0$ are $1/L$ and $J_2$. It follows that 
\begin{equation}
\label{eq:Scaling1}
f\left(L,J_2\right) = b^{-ld} f\left(b^{\lambda_1 l}\frac{1}{L},b^{\lambda_2 l}J_2\right)
\end{equation}
and by choosing $b^{\lambda_1 l}=L$ we get
\begin{equation}
\label{eq:Scaling2}
f\left(L,J_2\right) = \left(\frac{1}{L}\right)^{d/\lambda_1} f\left(1,\frac{J_2}{L^{-\lambda_2/\lambda_1}}\right)
\end{equation}
At $J_2=0$ this reduces to $f(L,0) = \mbox{const}\cdot L^{-d/\lambda_1}$ and because the ground-state energy per lattice site scales like $L^{-2}$, we conclude that $\lambda_1 =1$. \\
By setting  $b^{\lambda_2 l}=J_2^{-1}$ we find the relation
\begin{equation}
\label{eq:Scaling3}
f\left(L,J_2\right) = J_2^{d/\lambda_2} f\left(L^{-1}J_2^{-\lambda_1/\lambda_2},1\right).
\end{equation}
When inserting the known result $\lambda_2=3/2$, we can state that there must exist an universal scaling function $\Phi$ with
\begin{equation}
\label{eq:Scaling4}
J_2^{-4/3}\cdot f\left(L,J_2\right) = \Phi\left(L^{-1}J_2^{-2/3}\right) \; .
\end{equation}
We can do similar calculations for the gap $\Delta$ and get from $\Delta\left(L,J_2\right) = b^{-l} \Delta\left(b^{\lambda_1 l}L^{-1},b^{\lambda_2 l}J_2\right)$ the scaling relation
\begin{equation}
\label{eq:Scaling5}
J_2^{-2/3}\cdot \Delta\left(L,J_2\right) = \tilde{\Phi}\left(L\cdot J_2^{2/3}\right).
\end{equation}
To test this, we have applied the DMRG to 862 different chains with lengths up to 122 sites and $J_2 \in [0.00005,0.8]$ (see figure \ref{Skalenverhalten}). By these calculations the scaling relation (\ref{eq:Scaling5}) is confirmed in a convincing way. Note that practically all data points collapse on a one-dimensional manifold. Some minor deviations are noticeable and can be explained by higher order terms in the finite size scaling and an effective exponent 0.662 instead of 2/3 (see inset of fig. \ref{Skalenverhalten}). It is also possible to determine two scaling regions. For large lengths $L$ of the chain the plotted function saturates, indicating that the relation $\Delta E \propto J_2^{2/3}$, derived from scaling arguments for the infinite chain, really holds. For small $L \cdot J_2^{2/3}$ there is a linear regime, showing that the finite size gap proportional to $1/L$ is the dominant contribution in this region.

\section{Mapping onto a nl{\Large$\sigma$}-model}
\label{nonlinsigma}
In this section, we want to discuss the possible transition from the magnetically ordered phase to a disordered phase driven by the dimerization $\delta$ by using the nl$\sigma$-model as a low energy effective theory. The easiest way to get a path integral for the considered models is the use of coherent states.\cite{Klauder, Wiegmann, FradkinStone, Perelomov} Spin coherent states $\left|n\right>$ form an overcomplete basis set and are generated by a $SU(2)$-rotation of the highest weight state $\left|s,s\right>$ 
\begin{equation}
  \label{kohZustand}
  \left|\vec{n}\right> := \e^{\im\phi(\vec{n_0}\times \vec{n})\vec{S}} \left|s,s\right> \; ,
\end{equation}
where $\vec{n}_0$ is an unit vector along the quantization axis and $\cos(\phi) = \vec{n}\cdot\vec{n}_0$. By using the Trotter formula and inserting the identity operator, the partition function can be written as $Z=\int D\vec{n} \: \e^{-S_E[\vec{n}]}$ with an Euclidian action given by
\begin{equation}
  \label{Wirkung2}
  S_E[\vec{n}] = -\im s S_{WZ} + \int_0^\beta dt \left<\vec{n}\right| H \left|\vec{n}\right> \; .
\end{equation}
$S_{WZ}$ is a topological term (Wess-Zumino term), which arises from Berry phases and can be expressed as
\begin{equation}
  \label{Wirkung3}
S_{WZ}[\vec{n}] = \sum_{\vec{r}}\int_0^1 \! d\tau \int_0^\beta \! dt \: \vec{n}\cdot (\partial_t\vec{n}\times \partial_\tau\vec{n})
\end{equation}
with the boundary conditions $\vec{n}(t,0)=\vec{n}(t)$, $\vec{n}(t,1)=\vec{n_0}$ and $\vec{n}(0,\tau)=\vec{n}(\beta,\tau)$.

\subsection{$J_x/J_y$-model}
\begin{figure}[!ht]
\includegraphics [width=0.9\columnwidth]{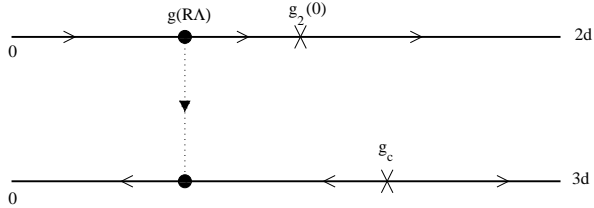}
\caption{RG flow for the $J_x/J_y$-model.}     
\label{RG-flow}
\end{figure}

With $a_0$ being the lattice constant, a site on the lattice $G$ can be described by $i = o + a_0 \sum_{a=1}^2 i_a\vec{e}^{\,a}$ with spatial unit vectors $\vec{e}^{\,a}$.
Using this notation, which is more suitable for the following calculations than (\ref{model-Ham}), the Hamiltonian of the model can be represented as
\begin{equation}
\label{Jx/Jy.1}
H = \sum_{i \in G} \{ J_x \vec{S}(i)\vec{S}(i+a_0\vec{e}^x) +  J_y \vec{S}(i)\vec{S}(i+a_0\vec{e}^y) \} \; .
\end{equation}
We assume periodic boundary conditions and an even number of sites $N$ in each direction. By defining $R=J_y/J_x$ and using the coherent state relation $\left<\vec{n}\right|\vec{S}\left|\vec{n}\right> = s\vec{n}$, the second term in (\ref{Wirkung2}) can now be easily calculated leading to the Euclidian action
\begin{eqnarray}
  \label{Jx/Jy.2}
  S_E[\vec{n}] &=& -\im s \: S_{WZ}[\vec{n}] \\
&&\!\!\!\!\!\!\!\!\!\!\!\!\!\!\!\!\!\!\!\!\!\!\!\!\!\!\! + J_x s^2 \sum_{i \in G} \int_0^\beta \!\!\! dt \:\{ \vec{n}(i)\vec{n}(i+a_0\vec{e}^x) + R \:\vec{n}(i)\vec{n}(i+a_0\vec{e}^y) \} \; . \nonumber
\end{eqnarray}
Now we use the well-known ansatz\cite{Polyakov, Haldane1}  
\begin{equation}
  \label{Ansatz}
  \vec{n}(i)=p(i)\vec{m}(i)\sqrt{1-a_0^{2d}\vec{l}^2(i)}+a_0^d\vec{l}(i) 
\end{equation}
with
\begin{equation}
  \label{factor_p}
p(i) = (-1)^{\sum_{a=1}^2 i_a} \; ,
\end{equation}
taking into account the short-range N\'eel order due to the antiferromagnetic exchange. Here $\vec{m}$ is the order parameter field and $\vec{l}$ represents the rapidly varying but small part. The constraint $\vec{n}^2=1$ leads to $\vec{m}^2=1$ and $\vec{m}\cdot\vec{l}=0$ and we will expand (\ref{Ansatz}) up to quadratic order. Because the Wess-Zumino term is independent of the microscopic details of the spin model, we want to discuss this term more generally in d dimensions. Starting with (\ref{Wirkung3}) and using the ansatz (\ref{Ansatz}) leads to  
\begin{eqnarray}
  \label{Wirkung4}
\im s S_{WZ}[\vec{n}] &=& is \int_V d^dx \int_0^\beta \! dt \: \vec{l}\cdot (\vec{m}\times \partial_t\vec{m}) \\
&+& \im \theta \sum_{i_2,\cdots,i_d=1}^N (-1)^{i_2+\cdots+i_d} \: k(i_2,\cdots,i_d), \nonumber 
\end{eqnarray}
where $\theta = 2\pi s$ and $k(i_2,\cdots,i_d)$ is the winding number or Pontryagin index of the field $\vec{m}$ defined by 
\begin{equation}
\label{Pontryagin}
k(i_2,\cdots,i_d) = \frac{1}{4\pi}\int dx_1 \int dt \;\partial_{x_1} \vec{m}\cdot (\vec{m}\times \partial_t \vec{m}).
\end{equation}
In one spatial dimension the second term in (\ref{Wirkung4}) is responsible for the different physics of chains with integer and half-integer spin. If the $\vec{m}$-field is smooth, the integer-valued Pontryagin index $k(i_2,\cdots,i_d)$ must be a constant and hence this term cancels out in higher dimensions.\cite{Haldane2} Note that we have to treat this term more carefully for the anisotropic models considered here, because the $\vec{m}$-field may no longer be smooth in each direction. Using the same ansatz (\ref{Ansatz}) for the part of the action depending on the Hamiltonian of the system and integrating out the $\vec{l}$-field, results in an effective action for the low lying excitations
\begin{equation}
  \label{Jx/Jy.3}
S[\vec{m}]=  \frac{\rho_s}{2} \!\! \int_V \!\!\! d^2x \!\int_0^\beta \!\!\!\! dt \left\{\! (\partial_x \vec{m})^2 + R (\partial_y \vec{m})^2 + \frac{1}{v_s^2} (\partial_t \vec{m})^2 \!\right\} 
\end{equation} 
with a spin stiffness $\rho_s = J_x s^2$, a transversal magnetic susceptibility  $\chi_\perp = 4 J_x a_0^2 (1+R)$ and a spinwave velocity $v_s = \sqrt{\rho_s \chi_\bot}$. We now rescale the imaginary time by $x_0 = v_s \cdot t$ leading to
\begin{equation}
  \label{Jx/Jy.4}
  S_{nl\sigma}[\vec{m}] = \frac{1}{2 a_0 g_0} \!\!\int \!\!d^3x \left\{ (\partial_{x_0} \vec{m})^2 \!+  (\partial_x \vec{m})^2 \!+ R (\partial_y \vec{m})^2 \right\}, \end{equation}
where the dimensionless coupling $g_0$ is defined by 
\begin{equation}
  \label{Jx/Jy.5}
   g_0 = \frac{2}{s}\sqrt{1+R} \; .
\end{equation}
We want to discuss (\ref{Jx/Jy.4}) following some arguments given by Affleck and Halperin.\cite{AffleckHalperin} Because the coupling in y-direction may be arbitrarily weak, a continuum representation may not be justified and we therefore rewrite the action as
\begin{widetext}
\begin{equation}
  \label{Jx/Jy.6}
  S_{nl\sigma}[\vec{m}] = \frac{1}{2g_0} \sum_n \int d^2x \: \left\{ (\partial_{x_0} \vec{m_n})^2 +  (\partial_x \vec{m_n})^2 + \frac{R}{a_0^2} (m_{n+1}-m_n)^2 \right\} .
\end{equation}
\end{widetext}
After a rescaling $y'=y/\sqrt{R}$ the momentum space UV-cutoff in y-direction is now smaller than the cutoff in the other directions. In a Wilsonian RG step, where the higher momentum modes are integrated out, only $k_0$ and $k_x$ contribute, so that the RG is essentially two-dimensional. Only if the momentum scale has been lowered so that also $k_y$ has components in the shell, we have to switch to the three-dimensional RG. During the two-dimensional RG, we also have to consider the rescaling of the $\vec{m}$-field which is given by $\vec{m}_n \rightarrow (\Lambda'/\Lambda)^x \vec{m}_n$. Here $\Lambda$ is the UV-cutoff before renormalization and $\Lambda'$ the reduced one after a RG step. $x$ is the scaling dimension of the $\vec{m}$-field which is equal $1/2$ for the Heisenberg model. \\
The scale $\Lambda'$, where we have to switch from two- to three-dimensional RG, is therefore given by the condition
\begin{equation}
  \label{Jx/Jy.7}
  R (\Lambda'/\Lambda)^{2x} \Lambda^2 /2g_0 \approx {\Lambda'}^2/2g(\Lambda') \; .
\end{equation}
Here $g(\Lambda')$ is the renormalized coupling constant when the momentum modes have been integrated out down to the cutoff $\Lambda'$. Note that the assumption that the $\vec{m}$-field is smooth on the scale of the lattice spacing is no longer justified in y-direction. Instead of canceling out, the second part of (\ref{Wirkung4}) leads to an independent winding number for each chain with topological angle $\theta = \pi$. Therefore $g$ flows in two-dimensions to the marginally stable fixed point $g_2(0)$ of the $s=1/2$ chain. As a consequence $g_0/g(\Lambda')$ is always of order one and (\ref{Jx/Jy.7}) simplifies to $\Lambda' = R \Lambda$. The coupling constant $g(\Lambda')$ then acts as bare coupling constant for the three-dimensional RG flow. Therefore the ground state is ordered if $g(\Lambda')$ is smaller than the critical fixed point $g_c$ of the three-dimensional RG, whereas it is disordered if $g(\Lambda') > g_c$. If even $g_2(0) < g_c$, the system is always ordered for a nonvanishing $R$. Numerical calculations using different methods \cite{SakaiTakahashi,KimBirgeneau,Affleck} give strong evidence that this model orders for arbitrarily weak $R$, meaning in the language of RG that $g_2(0)$ seems to be smaller than $g_c$ (see the corresponding flow diagram fig.\ref{RG-flow}).  

\subsection{stair-model}
We generalize the stair-model to d dimensions assuming a hypercubic, bipartite lattice $G = A \oplus B$ with periodic boundary conditions and an even number $N$ of sites in each spatial direction. We do this, because the model reduces in one dimension to a dimerized chain and we want to compare the results, especially the topological terms, for the dimerized chain and for the two-dimensional model. Using the same notation as for the $J_x/J_y$-model with the sum in (\ref{factor_p}) now running up to $d$, the Hamiltonian can be expressed by 
\begin{equation}
\label{eq}
H = J\sum_{i\in G}\sum_{a=1}^d \left[1+p(i)\delta \right] \vec{S}(i)\vec{S}(i+a_0\vec{e}^a) \, .
\end{equation}
Using again relation (\ref{Wirkung2}) and also the ansatz (\ref{Ansatz}), an Euclidian action for this model depending on the unit vector field $\vec{m}$ and the orthogonal vector field $\vec{l}$ is derived. By the same arguments given in the chapter before, the Wess-Zumino term vanishes in spatial dimensions greater than one under the assumption that $\vec{m}$ is smooth on the scale of the lattice spacing, but is important in one dimension. Again, we integrate out the rapidly varying but small $\vec{l}$-fields. The result is an effective action 
\begin{equation}
  \label{eq:Seff2}
  S[\vec{m}] =  \frac{\rho_s}{2} \int_V  \! d^dx \int_0^\beta \!\! dt \:\left[\gamma_s^{ab} (\partial_a \vec{m})(\partial_b \vec{m}) + \frac{1}{v_s^2}(\partial_t \vec{m})^2\right] 
\end{equation}
with a spin stiffness $\rho_s = Js^2 a_0^{2-d}(1-\frac{\delta^2}{d})$, an induced anisotropy $\gamma^{ab}_s = \delta^{ab}-\frac{\delta^2}{d-\delta^2}(1-\delta^{ab})$, a transversal magnetic susceptibility $\chi_\perp = 4Jd a_0^d$ and a spinwave velocity $v_s = \sqrt{\rho_s \chi_\perp}$. But there arises also an imaginary contribution proportional to $\delta$ from the $\vec{l}$-field integration, which can be expressed as
\begin{eqnarray}
\label{deltatopTerm}
&&\frac{\im \theta \delta}{d} \sum_{i_2,\cdots,i_d} \int dx_1 \int dt \;\partial_{x_1} \vec{m}\cdot (\vec{m}\times \partial_t \vec{m}) + \cdots \nonumber \\
&+& \frac{\im \theta \delta}{d} \sum_{i_1,\cdots,i_{d-1}} \int dx_d \int dt \;\partial_{x_d} \vec{m}\cdot (\vec{m}\times \partial_t \vec{m}) \\
&=& \frac{\im\theta \delta}{d} \sum_{i_2,\cdots,i_d} k_1(i_2,\cdots,i_d) + \cdots \nonumber \\
&+& \frac{\im\theta \delta}{d} \sum_{i_1,\cdots,i_{d-1}}  k_d(i_1,\cdots,i_{d-1}) \nonumber \, .
\end{eqnarray}
If the $\vec{m}$-field is smooth, every $k_i$ must be a constant. Under this assumption the topological term simplifies to
\begin{equation}
\label{deltatopTerm2}
S_{top} = \frac{\im\theta \delta}{d}N^{d-1}\sum_{a=1}^d k_a.
\end{equation}
In one dimension, where the stair-model corresponds to the dimerized chain, we find a total topological contribution of $S_{top}=\im \theta (1+\delta)k$ as has been calculated before.\cite{Affleck85} Before we discuss the two topological terms (\ref{Wirkung4}) and (\ref{deltatopTerm}) in two dimensions, we look at the additional anisotropy in the action (\ref{eq:Seff2}) expressed by the matrix $\gamma^{ab}_s$. This matrix is symmetric and becomes the identity if $\delta$ goes to zero. \\
In two dimensions a $45^o$-rotation diagonalizes this matrix and with $x_0 = v_s \cdot t$ the action of the two dimensional model is given by
\begin{equation}
  \label{Sd2}
  S_{eff}[\vec{m}] = \frac{1}{2}\int_\Omega d^3x \;\frac{\rho_s^a}{v_s}(\partial_a \vec{m})^2 \; .
\end{equation}
There are now different spin stiffnesses in the spatial directions given by 
\begin{equation}
  \label{Spinsteifigkeiten}
  \rho_s^1 = Js^2 = \rho_s \cdot \frac{2}{2-\delta^2} \quad ; \quad  \rho_s^2 = \rho_s \cdot \frac{2-2\delta^2}{2-\delta^2}
 % Js^2\cdot (1-\delta^2)=%
\end{equation}
and $\rho_s^0 = \rho_s$. By a rescaling of the imaginary time, (\ref{Sd2}) is transformed into 
\begin{equation}
  \label{Sd2.1}
  S [\vec{m}] \!=\! \frac{1}{2 a_0 g_0}\!\int \!\!d^3\tilde{x} \!\left\{ \!(\partial_{\tilde{x}} \vec{m})^2 \!+\! (1\!-\!\delta^2)(\partial_{\tilde{y}} \vec{m})^2 \!+\! (\partial_{x_0} \vec{m})^2 \right\}
\end{equation}
with a bare coupling given by
\begin{equation}
  \label{g0}
  g_0 = \frac{2 \sqrt{2}}{s} \; .
\end{equation}
Again the coupling in $\tilde{y}$ direction may be arbitrarily weak, and we therefore have to use a discrete version of (\ref{Sd2.1}) like for the $J_x/J_y$-model. As a consequence  a two-dimensional RG has to be used until the UV-cutoff is lowered to $\Lambda' = (1-\delta^2)\Lambda$. The coupling $g(\Lambda')$ is then the bare coupling for the three-dimensional RG. Now, we have to remember that there are also two topological terms (\ref{Wirkung4} and \ref{deltatopTerm2}) present. Because we have stated that it is necessary to use a discrete representation instead of derivatives for the weak couplings, neither the winding number in x- nor the winding number in y-direction is well defined any longer. We therefore have performed an alternative mapping to a nl$\sigma$-model starting with a slightly modified version of this model, where the chains formed by strong bonds are smooth and along the x-axis and the weak bonds form zigzag-chains (see figure \ref{stair-chains}). The result is again an anisotropic nl$\sigma$-model like (\ref{Jx/Jy.4}), but now without a topological contribution proportional to $\delta$ and with a winding number in the topological part of (\ref{Wirkung4}) calculated along the strong bonds. The situation is therefore exactly the same as in the $J_x/J_y$-model and if we accept $g_2(0) < g_c$ as an universal property, we conclude that the ground state of this model is always antiferromagnetically ordered for $\delta \in [0,1)$, and this order only vanishes at $\delta =1$, where the model consists of uncoupled critical chains.

\subsection{plaquette-model}
The model is described by the following Hamiltonian
\begin{eqnarray}
  \label{P.1}
  H &=& J \sum_{i,j} \{ \left[1+p_x(i) \delta \right] \vec{S}_{i,j}\vec{S}_{i+1,j} \nonumber \\
&+&  \left[ 1+p_y(i) \delta \right] \vec{S}_{i,j}\vec{S}_{i,j+1} \} \; ,
\end{eqnarray}
where we have defined $p_x(i)=(-1)^{i_x}$ and $p_y(i)=(-1)^{i_y}$. Substituting (\ref{P.1}) into (\ref{Wirkung2}), we derive again a path integral formulation. By using the same ansatz as before, we notice that the dimerization $\delta$ does only contribute in $(\partial_a \vec{m})^3 \cdot \vec{l}$ and higher orders. This is not surprising, because the relevant low energy modes are not only near wave vectors $(\pi,\pi)$ and $(0,0)$ but also near $(0,\pi)$ and $(\pi,0)$. We therefore generalize (\ref{Ansatz}) in the following way:
\begin{eqnarray}
  \label{P.4}
  \vec{n}(i)\!\!\! &=& \!\!\! p(i)\vec{m}(i)\sqrt{\!1-a_0^{2d}\big[\vec{l}_0(i)+p_x(i)\vec{l}_x(i)+p_y(i)\vec{l}_y(i)\big]^2} \nonumber \\
&& + a_0^d\big[\vec{l}_0(i)+p_x(i)\vec{l}_x(i)+p_y(i)\vec{l}_y(i)\big]
\end{eqnarray}
The constraint $\vec{n}^2=1$ implies $\vec{m}^2=1$ and $\vec{m}\cdot\vec{l}_i=0$, where $\vec{l}_i$ denotes any of the three fluctuation fields. By an expansion up to quadratic order in $\vec{l}_i$, $\partial_a \vec{m}$ and $\partial_t \vec{m}$, we get again the effective action (\ref{Jx/Jy.3}) but now $R=1$ and the other parameters are defined by $\rho_s = J s^2(1-\delta^2)$, $\chi_\perp = 8 J a_0^2$ and $v_s = \sqrt{\rho_s \chi_\bot}$. By rescaling the imaginary time axis this is transformed to the standard nl$\sigma$-model 
\begin{equation}
\label{standard-sigma}
S[\vec{m}] = \frac{1}{2a_0 g_0} \int_\Omega d^3x (\partial_\mu \vec{m})(\partial^\mu \vec{m})
\end{equation}
with a bare coupling
\begin{equation}
  \label{P.10}
  g_0 = \frac{2 \sqrt{2}}{s}\frac{1}{\sqrt{1-\delta^2}}. 
\end{equation}
All topological terms vanish for this model and - different from the two models considered before - the action is isotropic in the spatial directions. So the 3d RG has to be used from the beginning and whether or not the model has an ordered ground state depends on whether or not $g_0 < g_c$. Because $g_0 \rightarrow \infty$ for $\delta \rightarrow 1$, the order vanishes already before the plaquettes decouple. The precise value of $\delta_c \in (0,1)$ cannot be determined within the RG treatment.

\subsection{meander-model}
The Hamiltonian looks quite similar to (\ref{P.1}), but instead of an alternation with $p_x(i)$ in x-direction there is one with $p(i)$. As a result the $\vec{l}_y$-field is unnecessary and can be left out. By using this reduced form of (\ref{P.4}) and integrating out the $\vec{l}_0$- and $\vec{l}_x$-fields an effective theory is derived
\begin{widetext}
\begin{eqnarray}
  \label{meander.1}
S_{eff}[\vec{m}] &=& \frac{\im \theta}{2} \delta \sum_{i=1}^N k_x(i) \\
&+& \int_V \!\!d^2x \!\!\int_0^\beta \!\!dt \left\{ \frac{\rho_s}{2} \left[ (1-\frac{\delta^2}{2})(\partial_x \vec{m})^2 + (1-\delta^2) (\partial_y \vec{m})^2 +\frac{1}{v_s^2} (\partial_t \vec{m})^2 \right] \right\}, \nonumber 
\end{eqnarray}
\end{widetext}
where $\rho_s = J s^2$, $\chi_\perp = 8 J a_0^2$ and $v_s = \sqrt{\rho_s \chi_\bot}$. Like in the stair-model there are different spin stiffnesses in the spatial directions and also a topological contribution proportional to $\delta$ like (\ref{deltatopTerm2}), but in this model only the winding number in x-direction is involved. As in the stair-model, this winding number is not well defined. The same calculations within the modified model, where the chains formed by strong bonds are again smooth and along the x-axis (see figure \ref{meander-model}), show that the topological term proportional to $\delta$ vanishes and the winding number in (\ref{Wirkung4}) is calculated along the strong bonds. We therefore ignore the topological term in (\ref{meander.1}) from now on and use a 2d RG flow with $\theta=\pi$ later on. A rescaling of the imaginary time axis then leads to 
\begin{equation}
\label{meander.2} 
S[\vec{m}] \!=\! \frac{1}{2a_0 g_0} \!\!\int_\Omega \!\!\!d^3x \!\left\{ \!(\partial_x \vec{m})^2 \!+\! \frac{2-2\delta^2}{2-\delta^2}(\partial_y \vec{m})^2 \!+\! (\partial_t \vec{m})^2 \!\right\},
\end{equation}
where the bare coupling is given by
\begin{equation}
\label{meander.3}
g_0(\delta) = \frac{2\sqrt{2}}{s} \sqrt{\frac{2}{2-\delta^2}} \; .
\end{equation}
Because of the anisotropy in y-direction the RG is, like in the $J_x/J_y$- and stair-model, two-dimensional at the beginning. But here the bare coupling $g_0(\delta)$ is increased with increasing $\delta$. Therefore two scenarios are possible: If a $\delta_c$ exists so that $g_0(\delta) > g_2(0)$ if $\delta > \delta_c$, the coupling is driven by the 2d RG flow towards infinity and $g(\Lambda')$ with $\Lambda' = (2-2\delta^2)/(2-\delta^2)\Lambda$, which is the bare coupling for the 3d RG, is then greater than $g_c$. Therefore the ground state is disordered for $\delta>\delta_c$. If however even for $\delta \rightarrow 1$, $g_0(\delta)$ remains smaller than $g_2(0)$, the ground state is always ordered for a nonvanishing interchain coupling. In summary, we cannot decide within the RG treatment if an extended disordered phase exists or if there is long-range order for all $\delta \in [0,1)$ as in the $J_x/J_y$- and the stair-model.

\section{Linear spinwave theory (LSWT)}
\label{Linspinwave}
In the section before, we have mentioned that it is impossible to determine a value for the critical coupling $g_c$, or equivalent a value for the criticial dimerization $\delta_c$, for the plaquette- and the meander-model from RG. This is one reason to apply LSWT onto the considered models. The second reason is that we are interested in the question if these models can be the result of a dynamical process i.e.\!\! a spin-Peierls transition. Because the calculated ground-state energy for the isotropic two-dimensional Heisenberg antiferromagnet of $E_0/NJ \approx -0.6579$ deviates less than 3\% from the best numerical results \cite{BeardBirgeneau,Barnes} and also the sublattice magnetization $m= 0.3034$ agrees very well, we expect that LSWT gives reliable results near the isotropic point. On the other hand, the results for large dimerizations have to be regarded with care, because LSWT fails in one dimension. \\
We use the Holstein-Primakoff transformation \cite{HolsteinPrimakoff} to map the spin operators onto Bose operators:
\begin{equation}
  \label{H-P}
S^z=s-a^+a,\quad S^-=\sqrt{2s}\,a^+\sqrt{1-\frac{a^+a}{2s}}  
\end{equation} 
By expanding the square root in 1/s and taking only the lowest order into account, LSWT is reached.

\subsection{stair-model}
As before, we want to consider the stair-model generalized to d dimensions. Because the lattice $G=A \oplus B$ is bipartite, it is possible to write the Hamiltonian as 
\begin{eqnarray}
\label{eq:1}
H &=& \sum_{i\in A}\sum_{a=1}^{d}[J(1+\delta)\vec{S}(i)\vec{S}(i+a_0\vec{e}^a) \nonumber \\
&+& J(1-\delta)\vec{S}(i)\vec{S}(i-a_0\vec{e}^a)] \; ,
\end{eqnarray}
which is useful for LSWT. Starting point is again a N\'eel ordered state, and we therefore apply independent Holstein-Primakoff transformations to the two sublattices A, B:
\begin{subequations}
\begin{equation}
\label{eq:3}
x\in A: \: S^z(x) = s-a^+(x) a(x) \:\: ; \:\: S^-(x) = \sqrt{2s}\,a^+(x) 
\end{equation}
\begin{equation}
\label{eq:4}
x\in B: \: S^z(x) = -s+b^+(x) b(x) \:\: ; \:\: S^-(x) = \sqrt{2s}\,b(x)
\end{equation}
\end{subequations}
Taking the Fourier transform the Hamiltonian is bilinear and given by
\begin{eqnarray}
  \label{eq:12}
  H &=& -NJs^2d \\
&+& 2Jsd\sum_{\vec{k}}\left\{a^+_{\vec{k}} a_{\vec{k}} + b^+_{\vec{k}} b_{\vec{k}} + A a_{\vec{k}} b_{\vec{k}} +A^*a^+_{\vec{k}} b^+_{\vec{k}} \right\} \nonumber \; ,
\end{eqnarray}
 where the definitions $\gamma_{\vec{k}} = \frac{1}{d}\sum_{l=1}^d \cos(k_la_0)$, $\beta_{\vec{k}} = \frac{\im}{d}\sum_{m=1}^d \sin(k_ma_0)$ and $ A=\gamma_{\vec{k}}+\delta\beta_{\vec{k}}$ have been used. By means of a Bogoliubov transformation this is easily diagonalized leading to 
\begin{eqnarray}
\label{eq:Diag1}
H &=&  -NJds(s+1) \\
&+& 2Jsd\sum_{\vec{k}}\sqrt{1-\left(\gamma^2_{\vec{k}}-\delta^2\beta^2_{\vec{k}}\right)}\left(c^+_{\vec{k}} c_{\vec{k}} + d^+_{\vec{k}} d_{\vec{k}}+1\right) \; . \nonumber 
\end{eqnarray}

\subsubsection{Sublattice magnetization}
\begin{figure}[!ht]
\psfrag{d}{{\large $\displaystyle \delta$}}
\psfrag{m}{{\large $\displaystyle \langle S^z_x \rangle$}}
\psfrag{c}{{\large $\frac{1}{2}+\langle S^z_x \rangle-\Delta n_x$}}	
  \includegraphics [width=0.9\columnwidth]{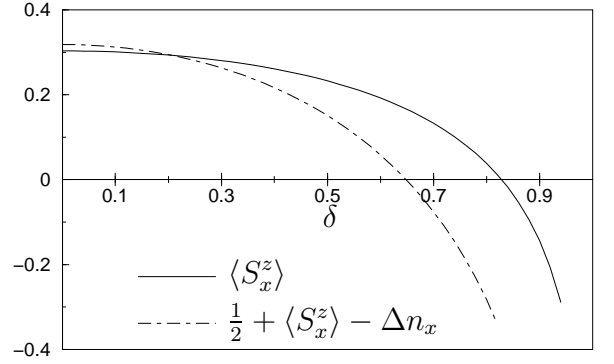}
 \caption{Sublattice magnetization and condition (\ref{TakaBed}) for the validity of the LSWT, both for the stair-model.}     
\label{Magnet_Treppen}
%% \setlength{\unitlength}{10mm}
%% \begin{picture}(0,0)
%% \put(4.2,4.05){\makebox(0,0){$\delta$}}
%% \put(-0.41,2.75){\makebox(0,0){$\left< S^z_x \right>$}}
%% \put(-0.2,2.1){\makebox(0,0){$\frac{1}{2}+\left< S^z_x \right>-\Delta n_x$}}
%% \end{picture}
  \end{figure}

The sublattice magnetization $m$ is given by
\begin{eqnarray}
\label{magn-stair}
m &=& \langle S^z_A\rangle =   \frac{N}{2}s - \sum_{\vec{k}} \langle a^+_{\vec{k}} a_{\vec{k}}\rangle \\
&=& \frac{N}{2}s - \frac{1}{2}\sum_{\vec{k}}\left\{\frac{1}{\sqrt{1-\left(\gamma^2_{\vec{k}}-\delta^2\beta_{\vec{k}}^2\right)}}-1 \right\} \nonumber
\end{eqnarray}
By replacing the sum by an integral this can be evaluated in principle in any dimension $d$. However in one dimension this integral is divergent. This is not astonishing due to Coleman's theorem \cite{Coleman} stating that the continuous $SU(2)$-symmetry cannot be broken in one dimension.  \\
In two dimensions the dimerization $\delta$ reduces the magnetization as expected and at $\delta=1$ there is again the one dimensional infrared divergence. Nevertheless, we can take the value of the dimerization where $m$ vanishes as indication for the breakdown of N\'eel order. For the stair-model we get $\delta_c = 0.8286$ or $R_c = (1-\delta)/(1+\delta) = 0.094$ (see figure \ref{Magnet_Treppen}). \\
To check the validity of the LSWT, we want to use an argument given by Sakai and Takahashi.\cite{SakaiTakahashi} In spinwave theory the number $n_x$ of bosons per lattice site is not restricted. That means that there are unphysical states in this theory, because in the original spin system the condition $n_x \leq 2S$ holds. A possible estimation for the validity of LSWT may therefore be given by
\begin{equation}
\label{TakaBed}
\langle n_x \rangle +\Delta n_x < 2s \;\Leftrightarrow \;\langle S^z_x \rangle - \Delta n_x > -s \; .  
\end{equation}
This predicts that LSWT is valid for this model from $\delta = 0$ up to $\delta \approx 0.65$ (see figure \ref{Magnet_Treppen}). Therefore the calculated $\delta_c$ is in a region, where LSWT is not reliable. We can only conclude that $\delta_c$ must be greater than $0.65$.

\subsubsection{Distortion due to spin-phonon coupling}
The dimerization leads to a gain in magnetic energy on the one hand, on the other hand it costs elastic energy. To answer the question if a spin-Peierls transition towards a distorted lattice is possible, we expand the calculated ground-state energy up to quadratic order. Because the linear term vanishes for symmetry reasons, we get
\begin{equation}
\label{groundenergy-Treppe}
\frac{E(\delta)}{NJ} = e_0 - A \cdot\delta^2
\end{equation}
with $e_0 = -0.6579$ and $ A = 0.261$. As mentioned in the introduction (\ref{spin-phonon}) the displacement of an atom is proportional to $\delta$ and the mean-field treatment of the phonons leads to a classical elastic energy proportional to $\delta^2$. Therefore the lattice is only distorted if 
\begin{equation}
\label{Dim-Bed}
\frac{m}{2} \frac{ J^2 \omega^2_{\mbox{\tiny trans}} (\pi,\pi)}{\lambda^2_{\mbox{\tiny trans}}(\pi,\pi)} < 0.261 \; ,
\end{equation}
whereas the lattice is unchanged otherwise. Here $m$ is the mass of the moved atom and $\omega$ the phonon frequency of the responsible phonon mode. If this condition is fulfilled, there is a coexistence of spin-Peierls dimerization and N\'eel-like long-range order. The situation here is quite different from that in one dimension. For the dimerized chain the magnetic energy scales like $E_{mag} \propto \delta^{4/3}$ and because the elastic energy is proportional to $\delta^2$, there is always a distortion. In the two-dimensional stair-model the magnetic and elastic energy behave like $\delta^2$ and therefore a distortion of the lattice is only possible if certain conditions are fulfilled. Because we expand around the isotropic point where LSWT results agree well with numerical results, the quantitative results for the condition of dimerization are expected to be reliable.

\subsection{$J_x/J_y$-model}
Because this model is well known and LSWT has already been applied to it,\cite{SakaiTakahashi} we only want to briefly state the results. The bosonic Hamiltonian is again easily diagonalized by a Bogoliubov transformation and can be written as
\begin{eqnarray}
  \label{J1/J2-Ham5}
& H &= -N(J_x+J_y)s(s+1)  \\ 
 &+& 2sJ_x(1+R)\sum_{\vec{k}}\gamma_{\vec{k}} \left(c^+_{\vec{k}}c_{\vec{k}}+d^+_{\vec{k}}d_{\vec{k}}+1\right) \nonumber
\end{eqnarray}
with $\gamma_{\vec{k}} = \sqrt{1-\left(\frac{1}{1+R}\right)^2 \left(\cos(k_x)+R\cos(k_y)\right)^2}$ and $R=J_y/J_x$. 

\subsubsection{Sublattice magnetization}
If we again calculate the sublattice magnetization and take its vanishing as indication for the breakdown of N\'eel order, we find a critical $R_c = 0.0337$. We also proved by using condition (\ref{TakaBed}) the reliability of the spinwave approach. It seems to be valid if $R>0.1$ and therefore we only can conclude that the critical coupling $R_c$ must be smaller than $0.1$. As mentioned before, numerical studies such as series expansions \cite{Affleck} and Monte Carlo studies \cite{KimBirgeneau} seem to indicate that $R_c$ is equal to zero.

\subsubsection{Distortion due to spin-phonon coupling}
By using the relation $\delta = (1-R)/(1+R)$ we switch again to the description of the model with the parameter $\delta$. In the expansion of the ground-state energy (\ref{groundenergy-Treppe}) the parameter $A$ is now given by $A = 0.146$. A spin-Peierls transition is only possible if the gain of magnetic energy is larger than the cost of elastic energy leading to the condition
\begin{equation}
\label{Dim-Bed2}
m \frac{J^2}{\lambda^2}\left(2 c^2_{\mbox{\tiny long}} (1,0)
+c^2_{\mbox{\tiny trans}}(1,1) -c^2_{\mbox{\tiny long}} (1,1) 
\right) < 0.146 \; ,
\end{equation}
where $c(1,0)$ and $c(1,1)$ are the phonon velocities in direction $(1,0)$
and $(1,1)$.

\subsection{plaquette-model}
The situation is more complicated for this model, because the unit cell includes four sites. Therefore we have to introduce four kinds of bosons. This happens as follows \\

\textbf{A:}  $r\in(2i,2j) $; $\;S^z=s-a_r^+a_r$;  $\; S^-=\sqrt{2s} \: a_r^+$ \\

\textbf{B:}  $r\in(2i+1,2j+1)$;  $\; S^z=s-b_r^+b_r$; $\; S^-=\sqrt{2s} \: b_r^+$ \\

\textbf{C:}  $r\in(2i,2j+1)$;  $\; S^z=-s+c_r^+c_r$;  $\; S^-=\sqrt{2s} \: c_r$ \\

\textbf{D:}  $r\in(2i+1,2j)$;  $\; S^z=-s+d_r^+d_r$;  $\; S^-=\sqrt{2s} \: d_r$ , \\

where A, B, C, D enumerate the four sublattices. \\
In principle it is possible to diagonalize every Hamiltonian of an assembly of $N$ bilinearly interacting bosons or fermions what is well-known since long.\cite{Bogoliubov} But especially for bosons it is often complicated to construct the transformation matrix $T$ between new and old operators, because the transformation is not unitary. We therefore add an ``antiferromagnetic field'' $B^z_A$, which allows us to calculate the sublattice magnetization without doing this canonical transformation explicitly. After taking the Fourier transform the Hamiltonian with additional field is given by
\begin{eqnarray}
  \label{Plaqu4}
  H \!\!&=&\!\! -2NJs^2 - N B_A^z s \nonumber \\
\!\!&+&\!\! 4Js\sum_{\vec{k}} \big\{ a_{\vec{k}}^+a_{\vec{k}}+b_{\vec{k}}^+b_{\vec{k}}+c_{\vec{k}}^+c_{\vec{k}}+d_{\vec{k}}^+d_{\vec{k}} \nonumber \\
\!\!&+& \!\! A_{\vec{k}}^x \left[a_{\vec{k}}c_{\vec{k}}+b_{\vec{k}}^+d_{\vec{k}}^+\right] \!+\! h.c. \!+\! A_{\vec{k}}^y \left[a_{\vec{k}}d_{\vec{k}}+b_{\vec{k}}^+c_{\vec{k}}^+\right] \!+\! h.c. \nonumber \\
\!\!&+&\!\! \frac{B_A^z}{4Js} \left[ a_{\vec{k}}^+a_{\vec{k}}+b_{\vec{k}}^+b_{\vec{k}}+c_{\vec{k}}^+c_{\vec{k}}+d_{\vec{k}}^+d_{\vec{k}}\right] \big\},
\end{eqnarray}
where $\gamma_{\vec{k}}^a = \frac{1}{2} \cos(\vec{k}a_0\vec{e}^a)$, $\beta_{\vec{k}}^a = \frac{\im}{2}\sin(\vec{k}a_0\vec{e}^a)$ and $A_{\vec{k}}^a = \gamma_{\vec{k}}^a+\delta\beta_{\vec{k}}^a$.\\
This has to be diagonalized under the subcondition that the new operators fulfill again Bose commutation relations leading to \cite{Tsallis} 
\begin{subequations}
\begin{equation}
  \label{Plaqu6}
  \mbox{(I)} \quad T^{-1}\mathcal{H}JT = \mathcal{H}_D J 
\end{equation}
and
\begin{equation}
  \label{Plaqu7}
  \mbox{(II)} \quad T^{+}JTJ = 1 \; ,
\end{equation} 
where $\mathcal{H}_D$ denotes the diagonalized Hamilton matrix and $J$ is given by
\begin{equation}
\label{Plaqu8}
J= \left(\begin{array}[c]{cccc}
-1&0&0&0 \\
0&-1&0&0 \\
0&0&1&0 \\
0&0&0&1 
\end{array}
\right) \, .
\end{equation}
\end{subequations}
From equation (II) it follows that $T$ is an element of the pseudounitary group $U(2,2)$, whereas (I) implies the secular equation
\begin{equation}
  \label{Plaqu9}
det(\mathcal{H}J-\lambda_i 1)=0  
\end{equation}
for the diagonalization problem. The eigenvalues of $\mathcal{H}$ are then given by $\left| \lambda_i \right|$. By solving the secular equation (\ref{Plaqu9}) we get the diagonalized Hamiltonian of the plaquette-model in LSWT:
\begin{subequations}
\begin{eqnarray}
\label{Plaqu12a}
H &=&  -2NJs(s+1) - N B_A^z(s+ \frac{1}{2}) \nonumber \\
&+& 4Js \sum_{\vec{k}} \big\{ \lambda_{\vec{k}}^1 \left[\mathcal{A}_{\vec{k}}^+\mathcal{A}_{\vec{k}}+\mathcal{C}_{\vec{k}}^+\mathcal{C}_{\vec{k}}+1\right] \nonumber \\
&+& \lambda_{\vec{k}}^2 \left[\mathcal{B}_{\vec{k}}^+\mathcal{B}_{\vec{k}} + \mathcal{D}_{\vec{k}}^+\mathcal{D}_{\vec{k}}+1\right] \big\} 
\end{eqnarray}
\begin{eqnarray}
\label{Plaqu12b}
\lambda_{\vec{k}}^1 &=& \sqrt{(1+\tilde{B}_A^z)^2-(|A_{\vec{k}}^x|+|A_{\vec{k}}^y|)^2} \nonumber \\
\lambda_{\vec{k}}^2 &=& \sqrt{(1+\tilde{B}_A^z)^2-(|A_{\vec{k}}^x|-|A_{\vec{k}}^y|)^2} 
\end{eqnarray}
\end{subequations}
with the new Bose operators $\mathcal{A}_{\vec{k}}$, $\mathcal{B}_{\vec{k}}$, $\mathcal{C}_{\vec{k}}$, $\mathcal{D}_{\vec{k}}$ and $\tilde{B}^z_A = B^z_A/4Js$. 

\subsubsection{Sublattice magnetization}
The ground-state sublattice magnetization per lattice site is now easily calculated from (\ref{Plaqu12a}) by the derivative $\frac{1}{N}\langle\frac{\partial H}{\partial B_A^z}\rangle\big|_{B_A^z=0}$. Setting $s=1/2$ and replacing the sum through an integral the sublattice magnetization is given by
\begin{equation}
\label{Plaqu13}
\langle S^z\rangle = 1 -\frac{1}{4 \pi^2} \int_{-\pi/2}^{\pi/2} \int_{-\pi/2}^{\pi/2} \left\{ \frac{1}{\tilde{\lambda}_{\vec{k}}^1} +  \frac{1}{\tilde{\lambda}_{\vec{k}}^2} \right\} \; ,
\end{equation}
where $\tilde{\lambda}_{\vec{k}}^i = \lambda_{\vec{k}}^i (\tilde{B}^z_A = 0)$.
This decreases with increasing $\delta$ and vanishes at $\delta_c = 0.798$ or aquivalently $R_c = 0.112$. This value agrees with that given in a paper by Koga, Kumada and Kawakami.\cite{KogaKawakami1} The same authors have also used a series expansion \cite{KogaKawakami2} starting from uncoupled plaquettes to determine the critical coupling and get $\delta_c \approx 0.3$. This value is in good agreement with results from Monte-Carlo calculations, showing again that the results in LSWT at high dimerizations have to be considered with care, because of the unphysical states in this approach. 

\subsubsection{Distortion due to spin-phonon coupling}
We set $\tilde{B}^z_A = 0$ and expand again the ground-state energy up to quadratic order. For this model the parameter $A$ is equal $0.174$. Therefore the lattice is distorted if the condition 
\begin{equation}
\label{Dim-Bed3}
\frac{m}{4}\left( \frac{J^2 \omega^2_{\mbox{\tiny long}} (\pi,0)}{\lambda^2_{\mbox{\tiny long}} (\pi,0)} + \frac{J^2 \omega^2_{\mbox{\tiny long}} (0,\pi)}{\lambda^2_{\mbox{\tiny long}} (0,\pi)}  \right) < 0.174
\end{equation}
is fulfilled.

\subsection{meander-model}
The sublattice structure of this model is similar to that of the plaquette-model and it is again necessary to introduce four kinds of bosons. With the same definitions as before the Hamiltonian of the model is given by
\begin{eqnarray}
  \label{Mod2.2}
 \!\!\!\!\!\!\! H &\!\! =\! &\!\! -2NJs^2 - N B_A^z s \nonumber \\
&+& 4Js\sum_{\vec{k}} \big\{ a_{\vec{k}}^+a_{\vec{k}}+b_{\vec{k}}^+b_{\vec{k}}+c_{\vec{k}}^+c_{\vec{k}}+d_{\vec{k}}^+d_{\vec{k}} \nonumber \\
&+& A_{\vec{k}}^x \left[a_{\vec{k}}c_{\vec{k}}+b_{\vec{k}}d_{\vec{k}}\right] + h.c. +A_{\vec{k}}^y \left[a_{\vec{k}}d_{\vec{k}}+b_{\vec{k}}^+ c_{\vec{k}}^+ \right] +h.c. \nonumber \\
&+& \frac{B_A^z}{4Js} \left[ a_{\vec{k}}^+a_{\vec{k}}+b_{\vec{k}}^+b_{\vec{k}}+c_{\vec{k}}^+c_{\vec{k}}+d_{\vec{k}}^+d_{\vec{k}}\right] \big\}.
\end{eqnarray}
By solving the secular equation (\ref{Plaqu9}) this is diagonalized leading to a Hamiltonian in the new Bose operators $\mathcal{A}_{\vec{k}}$, $\mathcal{B}_{\vec{k}}$, $\mathcal{C}_{\vec{k}}$, $\mathcal{D}_{\vec{k}}$ like (\ref{Plaqu12a}) but now with eigenvalues 
\begin{eqnarray}
  \label{eq:Plaqu11.1}
\!\!\!\lambda_{\vec{k}}^1 \!\!\!&=&\!\!\! \sqrt{\left(1+\tilde{B}_A^z\right)^2-\left|A_{\vec{k}}^x\right|^2 - \left|A_{\vec{k}}^y\right|^2 - 2\left|A_{\vec{k}}^y\right|\Re(A_{\vec{k}}^x)} \\
\!\!\!\lambda_{\vec{k}}^2 \!\!\!&=&\!\!\! \sqrt{\left(1+\tilde{B}_A^z\right)^2-\left|A_{\vec{k}}^x\right|^2 - \left|A_{\vec{k}}^y\right|^2 + 2\left|A_{\vec{k}}^y\right|\Re(A_{\vec{k}}^x)} \nonumber \; .
\end{eqnarray}

\subsubsection{Sublattice magnetization}
Analogous to the plaquette model the sublattice magnetization can be calculated by a derivative. The value for the critical dimerization determined by this calculation is $\delta_c = 0.898$ or $R_c = 0.054$. This is like in all the other models in a region where the spinwave approach is no longer justified.

\subsubsection{Distortion due to spin-phonon coupling}
The parameter in the expansion is given by $A = 0.160$ and therefore the condition for the spin-Peierls transition is
\begin{equation}
\label{Dim-Bed4}
\frac{m}{4} \left( \frac{J^2 \omega^2 (\pi,\pi)}{\lambda^2 (\pi,\pi)} + \frac{J^2 \omega^2_{\mbox{\tiny long}} (0,\pi)}{\lambda^2_{\mbox{\tiny long}} (0,\pi)}  \right) < 0.160 \; ,
\end{equation}
where we have assumed that the phonon frequencies and spin-phonon coupling constants for the longitudinal and transversal $(\pi,\pi)$-phonon are identical.

\section{Conclusions}
\label{Conclusions}
By using simple scaling arguments we have shown that for the $J_x/J_y$-model and the stair-model, which consist of weakly coupled Heisenberg chains for large dimerizations, the interchain coupling is a relevant operator. Therefore only at $\delta = 1$ the ground state is of one-dimensional nature. Also in the meander-model the interchain coupling is relevant, but there is also a relevant intrachain coupling. By numerical calculations (DMRG + TMRG) we have shown in this more complicated case that the scaling behaviour for the intrachain coupling is predicted correctly by the simple scaling argument. 
We also showed that it is possible to map all four models onto a nl$\sigma$-model. By applying the known results for the RG flow in two and three dimensions we have concluded that the ground state of the $J_x/J_y$- and the stair-model is N\'eel-like ordered for all $\delta \in [0,1)$, whereas an extended disordered phase exists for the plaquette-model. However, it was not possible to determine the critical dimerization $\delta_c \in (0,1)$ for the plaquette-model within the RG treatment. For the meander-model the RG gave no unique result. A phase diagram like for the $J_x/J_y$- and stair-model, but also an extended disordered phase as in the plaquette-model are both possible scenarios. The second possibility seems to be more probable, as the intrachain coupling has scaling dimension $x=1/2$ and is therefore more relevant than the interchain coupling with $x = 1$. If we start from $\delta =1$ and reduce the dimerization, we might expect that there arises first a system consisting of weakly coupled 2-leg ladders. Because a 2-leg ladder with $s=1/2$ is a gapped system, a small coupling between these ladders can be treated within normal perturbation theory and does not change the global properties drastically. At lower dimerization the gap closes and the system orders antiferromagnetically. If this picture is correct, $\delta_c$ must be smaller than one. \\
To investigate this further and to determine a value for the critical dimerizations $\delta_c$, we also applied LSWT to the models. The following values for $\delta_c$ are derived:
\begin{center}
\begin{tabular}[c]{|l|c|}
\hline model & \quad $\delta_c$ \quad \\
\hline\hline $J_x/J_y$-model & \quad 0.935 \quad \\
\hline meander-model & \quad 0.898 \quad \\
\hline stair-model & \quad 0.829 \quad \\
\hline plaquette-model \quad \quad & \quad 0.798 \quad \\
\hline
\end{tabular}
\end{center}
All these values have to be regarded with great care, because LSWT allows also unphysical states and a simple argument has shown that there is a large contribution of these states at such high dimerizations. \\
This problem does not occur near $\delta = 0$, where LSWT gives very precise results. We therefore believe that the conditions for the spin-Peierls transition, which we have obtained within LSWT, are qualitatively and quantitatively useful. For all models we have found that elastic and magnetic energies scale like $\delta^2$ at small dimerizations. A phase transition leading to a coexistence of spin-Peierls dimerization and antiferromagnetic long-range order is therefore only possible for certain values of the microscopic coupling constants. From an expansion of the magnetic ground-state energy $E(\delta) = e_0 - A \cdot \delta^2$ we have got the following values for the parameter $A$: 
\begin{center}
\begin{tabular}[c]{|l|c|}
\hline model & \quad $A$ \quad \\
\hline\hline stair-model & \quad 0.261 \quad \\
\hline plaquette-model \quad \quad & \quad 0.174 \quad \\
\hline meander-model & \quad 0.160 \quad \\
\hline $J_x/J_y$-model & \quad 0.146 \quad \\
\hline
\end{tabular}
\end{center}
The gain of magnetic energy is therefore largest for a stair-like distortion of the lattice, what is caused by a transversal $(\pi,\pi)$-phonon. What kind of distortion is energetically prefered, depends also on the elastic energy, which in general is different for each model. However, if we assume that the elastic energy is equal for all models, we would conclude that the stair-like distortion is energetically prefered. This is in contradiction to a result by Tang and Hirsch,\cite{TangHirsch} who have studied the plaquette- and stair-like distortion by an exact diagonalization of a $4\times 4$-lattice and conclude by using the same assumption that the plaquette structure is prefered. However, the lattice they have considered is very small and they have not done any finite size scaling so that we believe our result is more reliable for the infinite lattice.

\begin{acknowledgments}
JS thanks R. Raupach for doing the TMRG calculations and also J. Gruneberg for valuable discussions. KH gratefully acknowledges financial support by the Stipendienfonds im Verband der chemischen Industrie, BMBF and the Studienstiftung des dt. Volkes.
\end{acknowledgments}

\end{document}